\documentclass[pra,superscriptaddress,longbibliography, twocolumn,amsmath,amssymb]{revtex4-1}	
\usepackage{graphicx}
\usepackage{dcolumn} 
\usepackage{hyperref}
\hypersetup{
    colorlinks=true,        
    linkcolor=blue,          
    citecolor=blue,        
    filecolor=magenta,      
    urlcolor=blue           
}
\newcommand{\Figref}[1]{Fig.\ref{#1}}
\newcommand{\Tabref}[1]{Table~\ref{#1}}

\newcommand{\citenamefont}{}
\newcommand{\bibnamefont}{}

\usepackage{color} 				
\usepackage[normalem]{ulem} 			

\begin{document}
\title{Focusing of Drive and Test Bunches in a Dielectric Waveguide Filled with Inhomogeneous Plasma}

\author{G.V. Sotnikov}\email{sotnikov@kipt.kharkov.ua; Gennadiy.Sotnikov@gmail.com}
\affiliation{NSC Kharkov Institute of Physics and Technology, 61108 Kharkov, Ukraine}
\author{P.I. Markov}
\affiliation{NSC Kharkov Institute of Physics and Technology, 61108 Kharkov, Ukraine}
\author{I.N. Onishchenko}
\affiliation{NSC Kharkov Institute of Physics and Technology, 61108 Kharkov, Ukraine}

\date{\today}

\begin{abstract}
  The paper presents the results of numerical PIC-simulation of accelerated and drive bunches dynamics in a dielectric waveguide filled with radially inhomogeneous plasma. The wakefield was excited by the electron bunch in a quartz (permittivity 3.75) dielectric tube with outer and inner diameters of 1.2 mm and 1.0 mm, respectively, which was nested into a cylindrical metallic waveguide. The drive bunch characteristics were chosen to be: 5 GeV for electron energy, 3 nC for the charge, 0.2 mm – the bunch length, 0.9 mm – the bunch diameter. The accelerated bunch had the same parameters, except for the charge, which was equal to 0.3 nC. The interior of the waveguide was filled with plasma having different transverse density profiles, viz., the density profile formed in the capillary discharge, and the radially nonuniform density profile with the vacuum channel along the waveguide axis. For all the cases under study the plasma density was low, so that the plasma frequency was lower than the fundamental dielectric mode frequency. The obtained PIC-simulation data have shown that the vacuum channel in the inhomogeneous plasma cylinder improves the accelerated bunch focusing. There is the optimum vacuum-channel size value, at which the focusing turns out to be the strongest. The improvement in the accelerated bunch focusing is accompanied by the decrease in the accelerating gradient as compared with the full plasma filling of the drift channel. The best acceleration takes place in the absence of plasma; however in that case the test bunch focusing does not occur.
\end{abstract}
\maketitle

\section{Introduction}
The dielectric wakefield accelerator (DWA) is a promising candidate for building the TeV-energy range electron-positron collider \cite{Shiltsev_2012,OShea2016, Thompson2008PRL, Gai1988PRL,Shchelkunov2012PRSTAB}. However, among the DWA’s shortcomings we should mention the susceptibility of bunches to the beam breakup instability \cite{Gai1997PRE,She2008TP} (BBU) peculiar to electron linear accelerators \cite{Bal1983ICHEA-12}. The excitation of higher-order modes \cite{Ng990PRD, Ros1990PRD} results in the limitation of the achievable accelerating-field value with an increase in the drive bunch charge \cite{Gai2014PRSTAB}. Besides, high charges cause the beam quality degradation.
To suppress the BBU instability and for focusing the driving and accelerated bunches, it has been recently suggested that the DWA drift channel should be filled with a plasma of certain density~\cite{Sot2014NIMA}; this structure has been called the plasma dielectric wakefield accelerator (PDWA). The focusing in the structure occurs owing to the excitation of the plasma wakefield. The distinctive feature of the PDWA that makes it attractive for colliders is the possibility of focusing both the electron and positron bunches ~\cite{Sot2014NIMA}. Analytical studies and numerical simulations have confirmed the possibility of focusing the driving and accelerated bunches at filling the drift channel with plasma not only for the cylindrical DWA configuration \cite{Kniaziev2015PAST-NPI, Kniaziev2016NIMA}, but for the rectangular one as well \cite{Mar2016PAST-NPI}. The PDWA studies \cite{Sot2014NIMA} were carried out for the plasma density, which is homogeneous in the transverse section of the drift channel, and can be generated by an external source. The other way of plasma creation, widely used in the studies of wakefield acceleration methods, is the capillary discharge \cite{Ehrlich1996PRL,Butler2002PRL,Spence2001PRE,PompiliArXiv2019}. In this case, at great times of the discharge onset, the nonuniform plasma density distribution is formed with the radial dependence obtained by numerical simulation of the discharge in ref. \cite{Bob2001PRE}, which can be described by the parabolic relation for a considerable part of the channel section \cite{Stein2006PRSTAB}. The PIC simulation of the test electron bunch acceleration with the model plasma-density dependences~\cite{Bob2001PRE,Stein2006PRSTAB} has shown that the capillary discharge plasma improves the focusing of accelerated bunches \cite{Sot2017EPJ} in comparison with the case of the uniform plasma density in the section. At present, in studies of the concept of the beam-driven plasma wakefield accelerator (PWFA), for improving the positron bunch transport it has been suggested that the hollow plasma channel should be used (see refs. \cite{Joshi_2018,Ges2016Ncom} and the references cited there). The hollow plasma channel, and in the limiting case the vacuum channel in plasma, aids in improving the transport of the electron bunch, too \cite{Bal1997PPR}. In the present study, we investigate by the PIC simulation the effect of the vacuum channel in the radially inhomogeneous plasma on focusing of the drive and accelerated electron bunches in the PDWA.

\section{The statement of the problem}
The wakefield structure under study represented a metallic cylindrical waveguide of radius $b$ , into which the dielectric tube of inner radius $a$  was inserted tightly. The transport channel (region inside the dielectric tube ${r_{p1}} \le r \le a$  ) was partially filled with plasma, having generally a nonuniform radial density profile.

The cylindrically-shaped drive electron bunch of radius   passed through the slow-wave structure along its axis and excited the wakefield. In a given delay time $t_{del}$ , following the drive bunch, the test bunch was injected into the system along its axis, the test bunch charge being $10$ times less than it was for the drive bunch. The delay time of the test bunch was chosen such that the bunch could get into the accelerating phase of the drive bunch wakefield.

\begin{figure}
  \centering
  \includegraphics[width=0.5\textwidth]{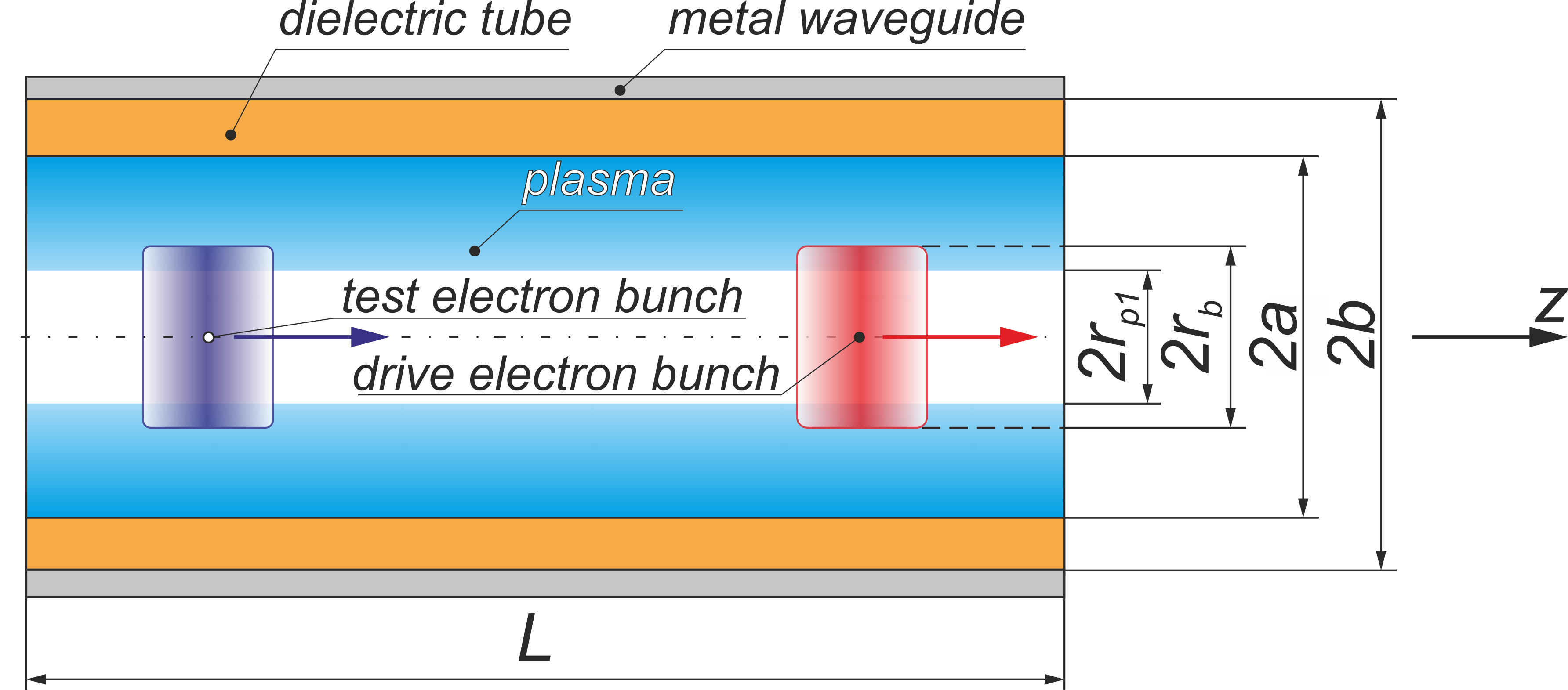}
  \caption{Schematic sketch of the longitudinal section of the PDWA. The pink cylinder shows the drive electron bunch, the violet cylinder – the accelerated (test) bunch. The plasma-filled region is depicted in blue, and the dielectric tube --- in orange.}\label{Fig:01}
\end{figure}

The initial sizes of the drive and accelerated bunches are equal. The structure with the drive and accelerated bunches is called here the plasma-dielectric wakefield accelerator (PDWA). Its schematic is shown in \Figref{Fig:01}.

\begin{ruledtabular}
\begin{table}
\centering
\caption{PDWA parameters used in the computations}\label{tab:01}
\begin{tabular}{|l|r|}
Inner dielectric-tube radius $a$             & $0.5\,mm$          \\
Outer dielectric-tube radius $b$             & $0.6\,mm$          \\
Inner plasma-cylinder radius $r_{p1}$        & $0\div 0.5\,mm$    \\
Dielectric permittivity $\varepsilon$        & $3.75 (quarz)$     \\
Bunch energy $E_0$                           & $5\,GeV$           \\
Drive bunch charge                           & $3\,nC$            \\
Test bunch charge                            & $0.3\,nC$          \\
Longitudinal rms deviation of bunch          & $0.1\,mm$          \\
charges $2\sigma$(Gauss charge distribution) &                    \\
Total bunch length used in PIC simulation    & $0.2\,mm$          \\
Bunch diameter $2r_b$                        & $0.899\,mm$        \\
Paraxial plasma density $r_{p1} = 0$         & $4.41\cdot 10^{14}\,cm^{-3}$ \\
\end{tabular}
\end{table}
\end{ruledtabular}

The initial parameters of the dielectric waveguide and drive bunch were chosen to be the same as in the studies of the wakefield excitation by the relativistic electron bunch in the PDWA with homogeneous plasma \cite{Sot2014NIMA}. The parameters provided the wakefield excitation of the THz frequency range (the dielectric-wave mode frequency  ${E_{01}}$ in the absence of plasma in the transport channel was equal to $\sim 297\,GHz$). \Tabref{tab:01} lists the parameter values of the waveguide, and also, of drive and test accelerated bunches used in numerical simulation.

\begin{figure}
  \centering
  \includegraphics[width=0.5\textwidth]{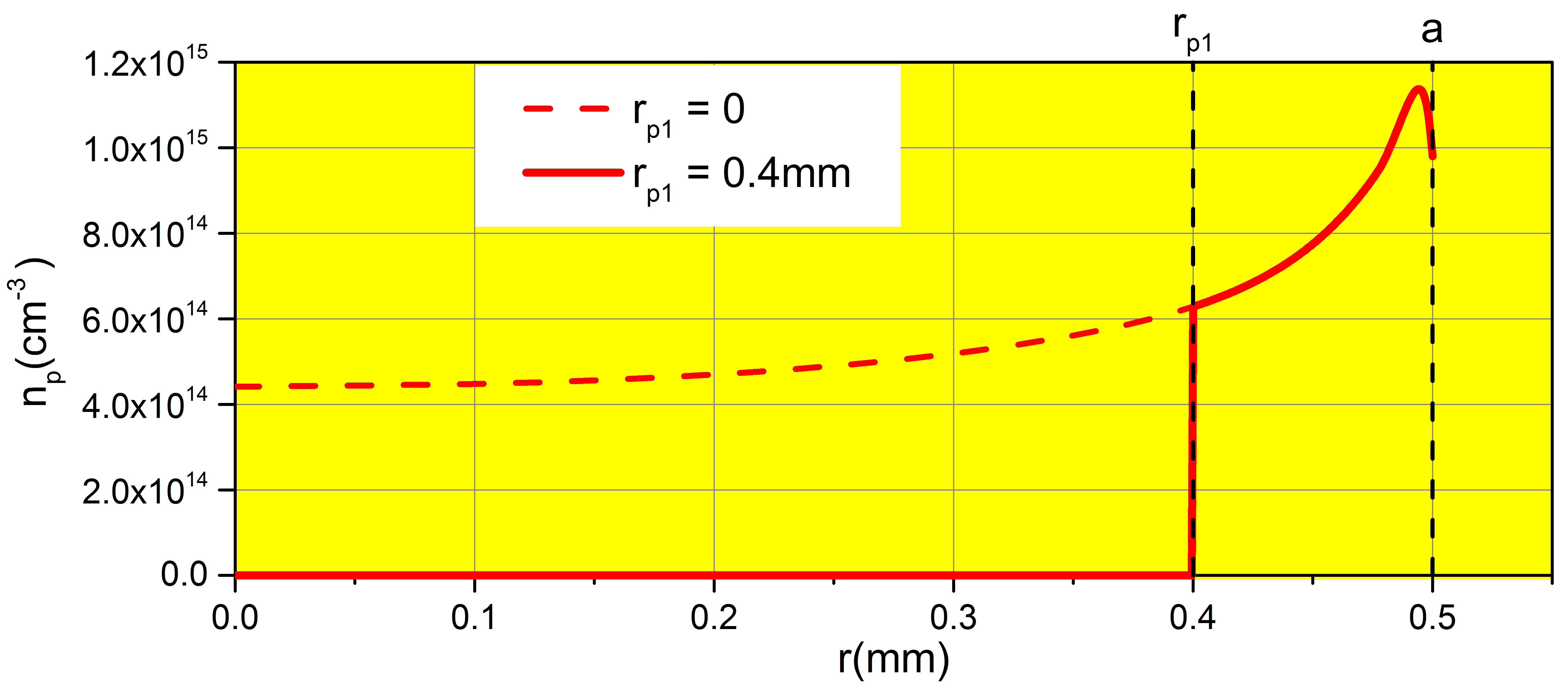}
  \caption{Model dependence of plasma density on the radius r for two cases: a) the inner region of the dielectric tube fully filled with plasma (dashed line), and b) the plasma layer with   (solid curve).  }\label{Fig:02}
\end{figure}

To describe the plasma density profile, we started from the profile of density decrease from the dielectric surface, generally realizable at the capillary discharge. It was determined numerically by N. A. Bobrova et al. \cite{Bob2001PRE}. At some point $r = {r_{p1}}$, the plasma density fell off to zero, forming the vacuum channel (see \Figref{Fig:02}). In all the computations, the plasma density, extrapolated to the structure axis with the parabolic relation \cite{Stein2006PRSTAB}, was put to be $4.41\cdot 10^{14}\,cm{-3}$, this corresponding to the plasma wave frequency of $\sim 189\,GHz$.

\section{Results of 2.5-dimensional PIC-simulation}

In numerical simulation by means of our 2.5D PIC-code we investigated the wakefield topography and dynamics of electron bunches in their motion in the drift chamber. The studies were performed for several variants with different inner plasma-cylinder radius   values varying in the range from $0$ to $0.5\,mm$.

\begin{figure*}[!tbh]
  \centering
  \includegraphics[width=\textwidth]{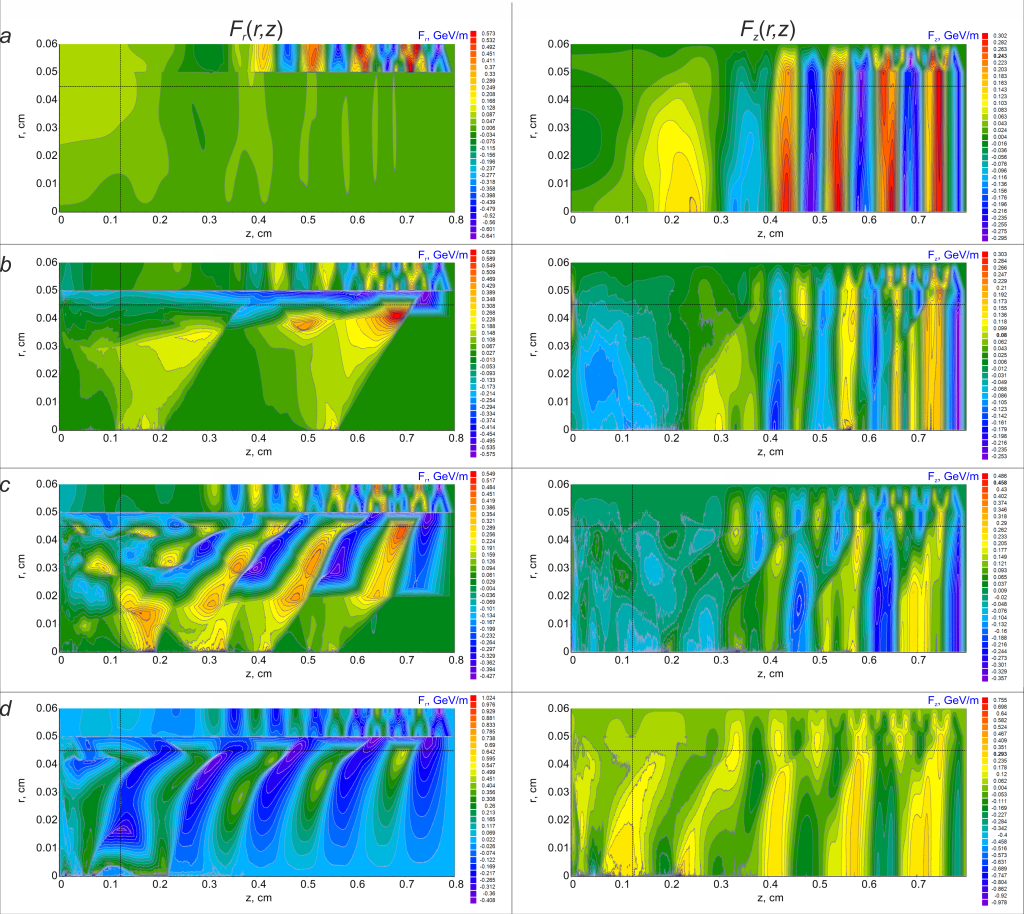}\hfill
  \caption{Color maps and level lines for the transverse  ${F_r}\left( {r,z} \right)$  (left) and longitudinal ${F_z}\left( {r,z} \right)$  (right) Lorentz force components acting on the test electron for the time  $t = 26.69\;{\rm{ps}}$ (the driving bunch has reached the structure end) at different ${r_{p1}}$   values: a) ${r_{p1}} = 0.5\;{\rm{mm}}$, b)${r_{p1}} = 0.4\;{\rm{mm}}$, c) ${r_{p1}} = 0.2\;{\rm{mm}}$, d) ${r_{p1}} = 0$}\label{Fig:03}
\end{figure*}

\Figref{Fig:03} shows comparative shots of the force components acting on the test electron in the PDWA for the time $t = 26.69\;{\rm{ps}}$  (the drive bunch has reached the structure end) at different ${r_{p1}}$  values: a) ${r_{p1}} = 0.5\;{\rm{mm}}$, b)${r_{p1}} = 0.4\;{\rm{mm}}$, c) ${r_{p1}} = 0.2\;{\rm{mm}}$, d) ${r_{p1}} = 0$. Note that the case a) corresponds to the absence of plasma in the drift region, and the case d) – to the complete plasma filling of the said region. From now on the data shown for the case d) have been partially presented in our paper \cite{Sot2017EPJ}. It can be seen that the cases b) to d) encourage one to choose the position of the accelerated bunch such as to accelerate and focus the bunch simultaneously.

\begin{figure*}
  \centering
  \includegraphics[width=\textwidth]{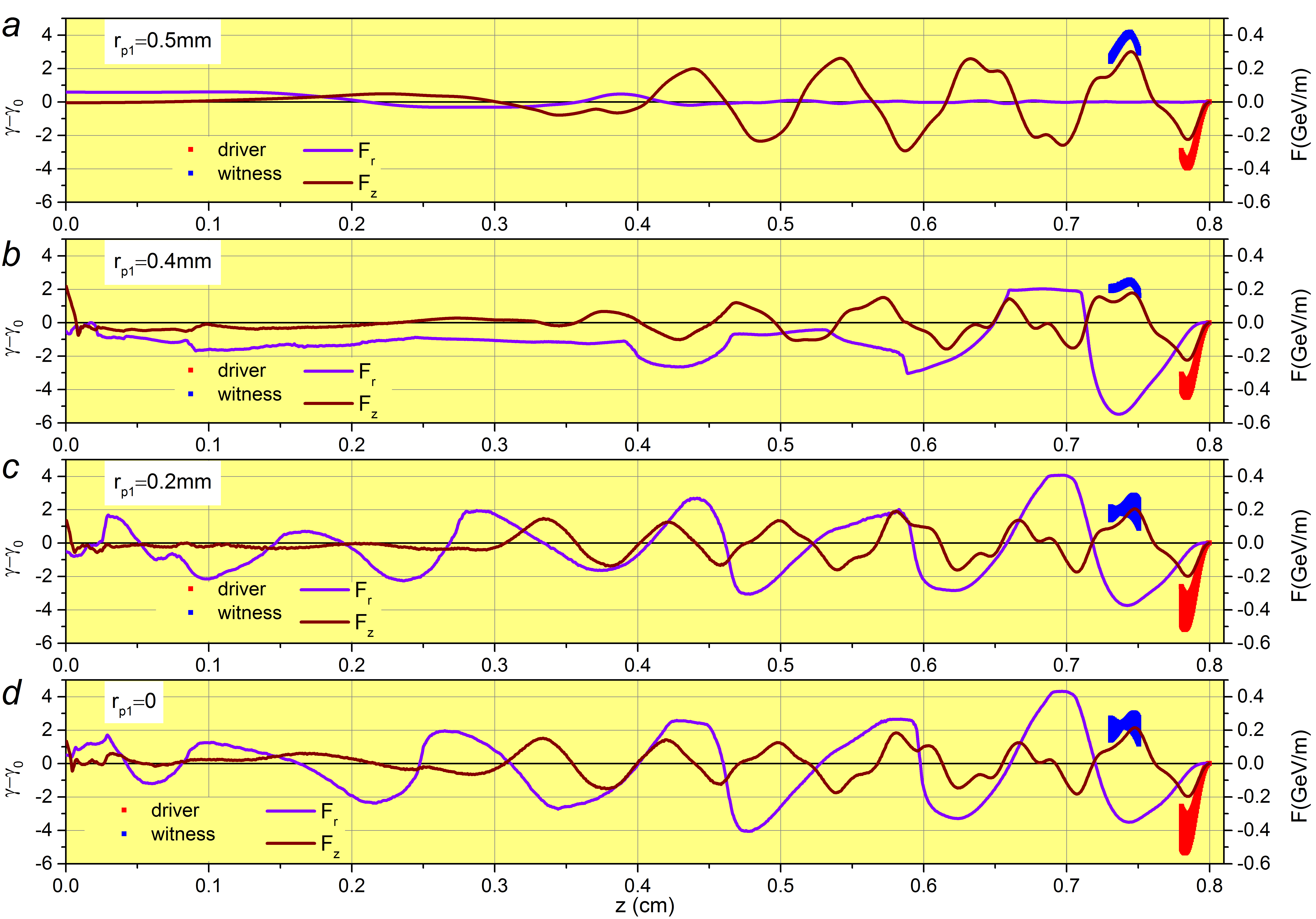}
  \caption{Energy – longitudinal coordinate phase space combined with the dependences of the longitudinal   ${F_z}\left( z \right)$ and transverse ${F_r}\left( z \right)$  forces at $r = 0.45\;{\rm{mm}}$  for the same instant of time and the same ${r_{p1}}$  values, as in \Figref{Fig:03}. The electron energies of the accelerated bunch and of the driver bunch are shown in blue and red, respectively.}\label{Fig:04}
\end{figure*}

To illustrate the mechanism of the accelerated bunch focusing, \Figref{Fig:04} shows the phase space combined with the relations of longitudinal ${F_z}\left( z \right)$ and transverse ${F_r}\left( z \right)$  forces at $r = 0.45\;{\rm{mm}}$  for the same time as given in \Figref{Fig:03} for different plasma density-radius dependencies. The red color in the phase space (see \Figref{Fig:04}) shows the electron energy of the drive bunch. The blue color shows the electron energies of the accelerated (test) bunch. The test bunch delay was chosen such as to provide the bunch acceleration in the first local maximum of the longitudinal force ${F_z}\left( z \right)$. For the cases b) to d) this delay also provided the transverse focusing by the transverse force ${F_r}\left( z \right)$.

As it follows from the plots shown in \Figref{Fig:03}a and \Figref{Fig:04}a, in the absence of plasma in the drift channel, the transverse force ${F_r}\left( z \right)$  for the ultrarelativistic drive bunch is practically equal to zero. As a consequence, the focusing of the test bunch is also absent. In the ${r_{p1}} = 0.4\;{\rm{mm}}$  case, the plasma layer thickness makes $0.1\,mm$, and the electron bunches blow over the plasma only slightly (over $0.05\,mm$). At that, as it follows from \Figref{Fig:04}b, the intense transverse force   gets excited. The local minimum of this force, equal to $–0.55\,GeV/m$, is observed in the region, where the test bunch is found, and this results in good focusing of the latter.

With a further decrease in ${r_{p1}}$, the plasma layer thickness also increases, and the increasing part of electrons of bunches are localized in plasma. \Figref{Fig:04}c and \Figref{Fig:04}d illustrate the cases with ${r_{p1}} = 0.2\;{\rm{mm}}$  and $r_{p1} = 0$, respectively. It can be seen that here the transverse focusing force ${F_r}\left( z \right)$  also gets excited, but its minimum in the position of the test bunch is somewhat smaller, viz., $-0.37\,GeV/m$ and $-0.35\,GeV/m$, respectively. As a result, the focusing of the bunch becomes weaker.

\begin{figure*}
  \centering
  \includegraphics[width=\textwidth]{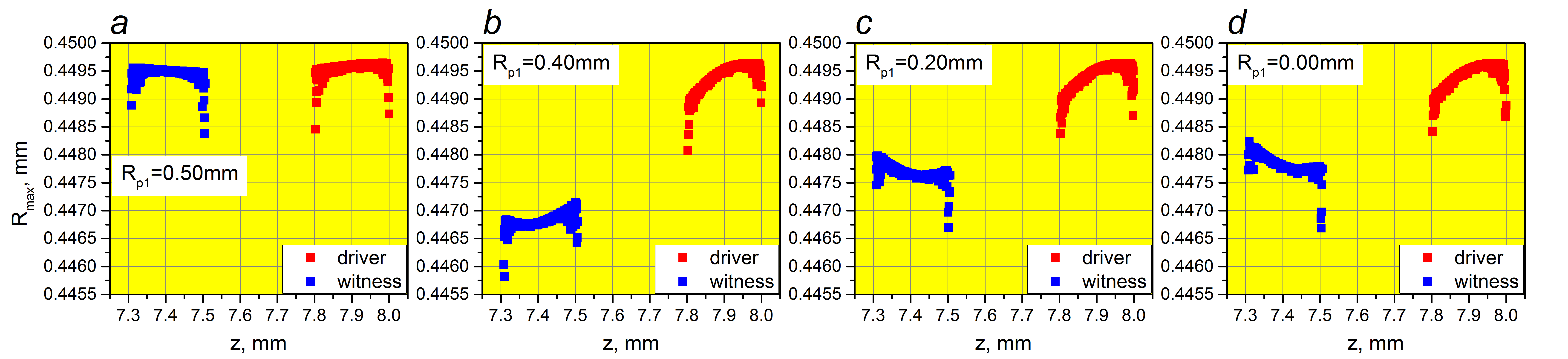}
  \caption{Configuration space representing the position of the bunch electrons being at the periphery for   at different ${r_{p1}}$   values: a) ${r_{p1}} = 0.5\;{\rm{mm}}$, b)${r_{p1}} = 0.4\;{\rm{mm}}$, c) ${r_{p1}} = 0.2\;{\rm{mm}}$, d) ${r_{p1}} = 0$.}\label{Fig:05}
\end{figure*}

To demonstrate the test bunch focusing for the above-mentioned four plasma density cases, \Figref{Fig:05} shows the configuration space representing the position of the drive and test bunch electrons being at the periphery (having the maximum radius at the given $z$, ${R_{\max }}\left( z \right)$ ) for the time $t = 26.69\;{\rm{ps}}$.
As it follows from the plots given in \Figref{Fig:05}, in the absence of plasma in the drift channel (case a),${r_{p1}} = 0.5\;{\rm{mm}}$  ), no focusing of the test bunch is observed. The case b) with  ${r_{p1}} = 0.4\;{\rm{mm}}$ exhibits the strongest focusing of the test bunch. In the cases c) at ${r_{p1}} = 0.2\;{\rm{mm}}$   and d) ${r_{p1}} = 0$  the test bunch focusing is weaker than in the case b). Note that in the case of full plasma filling of the drift channel (case d), the test bunch is focused worse than in the case of the vacuum channel presence (cases b) and c)).

It can be also seen in \Figref{Fig:05} that the drive bunch shows up the focusing only in its back part with the radius at the head part of the bunch remaining unchanged (see \Figref{Fig:05}b),c),and d)). This is in agreement with the transverse force dependence ${F_r}\left( z \right)$    given in \Figref{Fig:04}b),c) and d).

\begin{figure}
  \centering
  \includegraphics[width=0.47\textwidth]{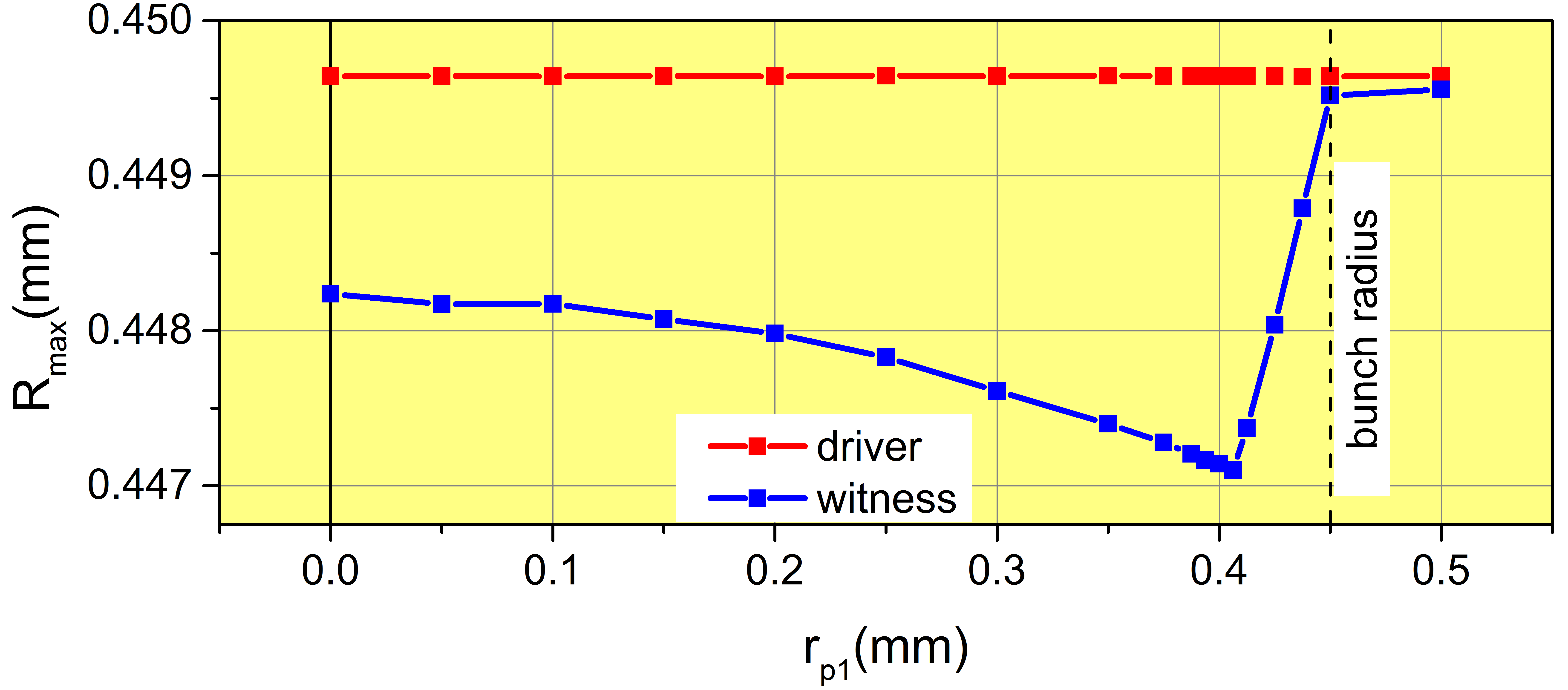}
  \caption{Bunch radius ${R_{\max }}$  versus the radius of the vacuum channel, ${r_{p1}}$ , for the time $t = 26.69\;{\rm{ps}}$.}\label{Fig:06}
\end{figure}

The behavior of the radii of the drive and accelerated bunches ${R_{\max }}$  with variation of the vacuum channel radius ${r_{p1}}$  from $0$ to $0.5\,mm$ for the time $t = 26.69\;{\rm{ps}}$  is shown in \Figref{Fig:06}. As it follows from the obtained dependences, the increase in the ${r_{p1}}$    from $0$ up to $0.40625\,mm$ causes the improvement in the test bunch focusing, whereas with a further increase in  ${r_{p1}}$  the focusing gets worse. If the bunches are entirely in the vacuum channel, i.e., the plasma layer is outside the bunches, the focusing is absent.

\begin{figure*}
  \centering
  \includegraphics[width=\textwidth]{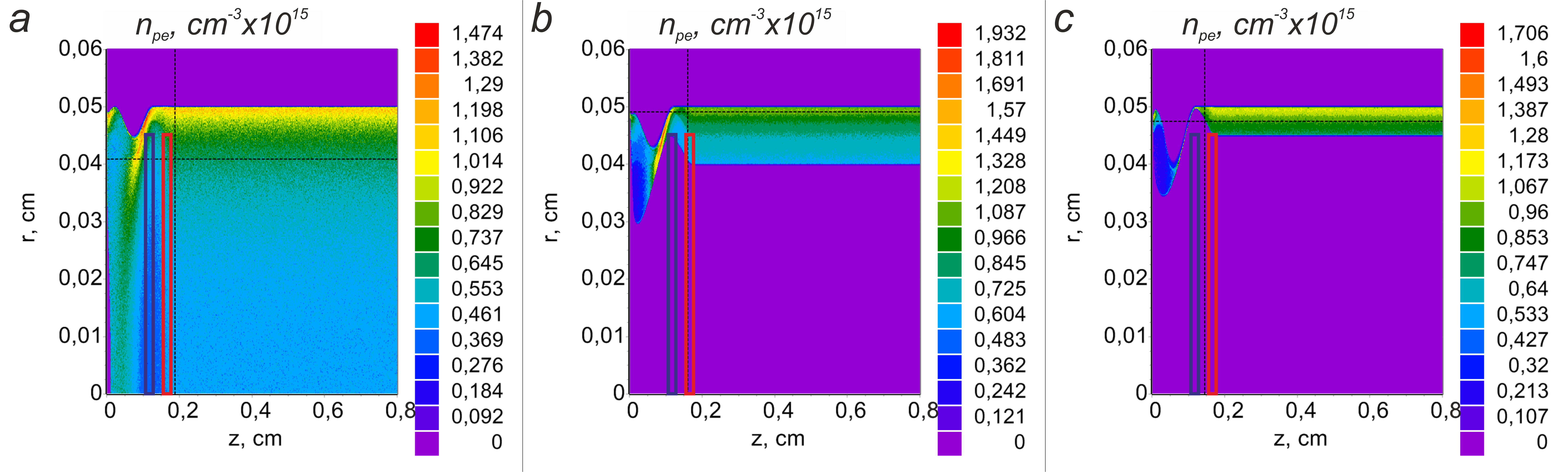}
  \caption{Plasma electrons density ${n_{pe}}\left( {r,\;z} \right)$  for  $t = 5.9\;{\rm{ps}}$ at ${r_{p1}} = 0$ (a),  ${r_{p1}} = 0.4\;{\rm{mm}}$ (b) and  ${r_{p1}} = 0.4\;{\rm{mm}}$ (c). The red and blue rectangles show the positions of the drive and test bunches, respectively.}\label{Fig:07}
\end{figure*}

To explain the behavior of the transverse size of the bunches, shown in \Figref{Fig:06}, let us analyze the behavior of the plasma electrons in the drift channel. \Figref{Fig:07} shows the plasma electron densities   ${n_{pe}}\left( {r,\;z} \right)$ for the time  $t = 5.9\;{\rm{ps}}$ at full plasma filling of the drift region, ${r_{p1}} = 0$(a), at ${r_{p1}} = 0.4\;{\rm{mm}}$  (b) and at ${r_{p1}} = 0.45\;{\rm{mm}}$  (c). The red and blue rectangles in \Figref{Fig:07} show the positions of the drive and test bunches, respectively.
In the case of full plasma filling of the drift region, at  ${r_{p1}} = 0$ (see \Figref{Fig:07}a), the above described processes lead to the formation of the region with a reduced plasma electrons density, and hence, to the excess of plasma ions in the area, where the test bunch is present. The excess of ions not only directs the plasma electrons to the structure axis, but also focuses the test bunch electrons being on all radii of the bunch.

At ${r_{p1}} = 0.4\;{\rm{mm}}$  (see  \Figref{Fig:07}b) in the area, where the test bunch is present, the plasma electrons are practically absent. Hence, the plasma ions focus the test bunch electrons being only on the radii from ${r_{p1}}$  to $r_b$ , i.e., from $0.4\,mm$ to $0.45\,mm$, this appearing, however, sufficient for attaining the maximum focusing of the bunch.

If the test bunch moves in the vacuum channel, i.e., when the plasma tube surrounds the electron bunches (see \Figref{Fig:07}c), the excess of plasma ions as compared to the plasma electrons is observed outside the test bunch, and this has no effect on its focusing.

\begin{figure}
  \centering
  \includegraphics[width=0.47\textwidth]{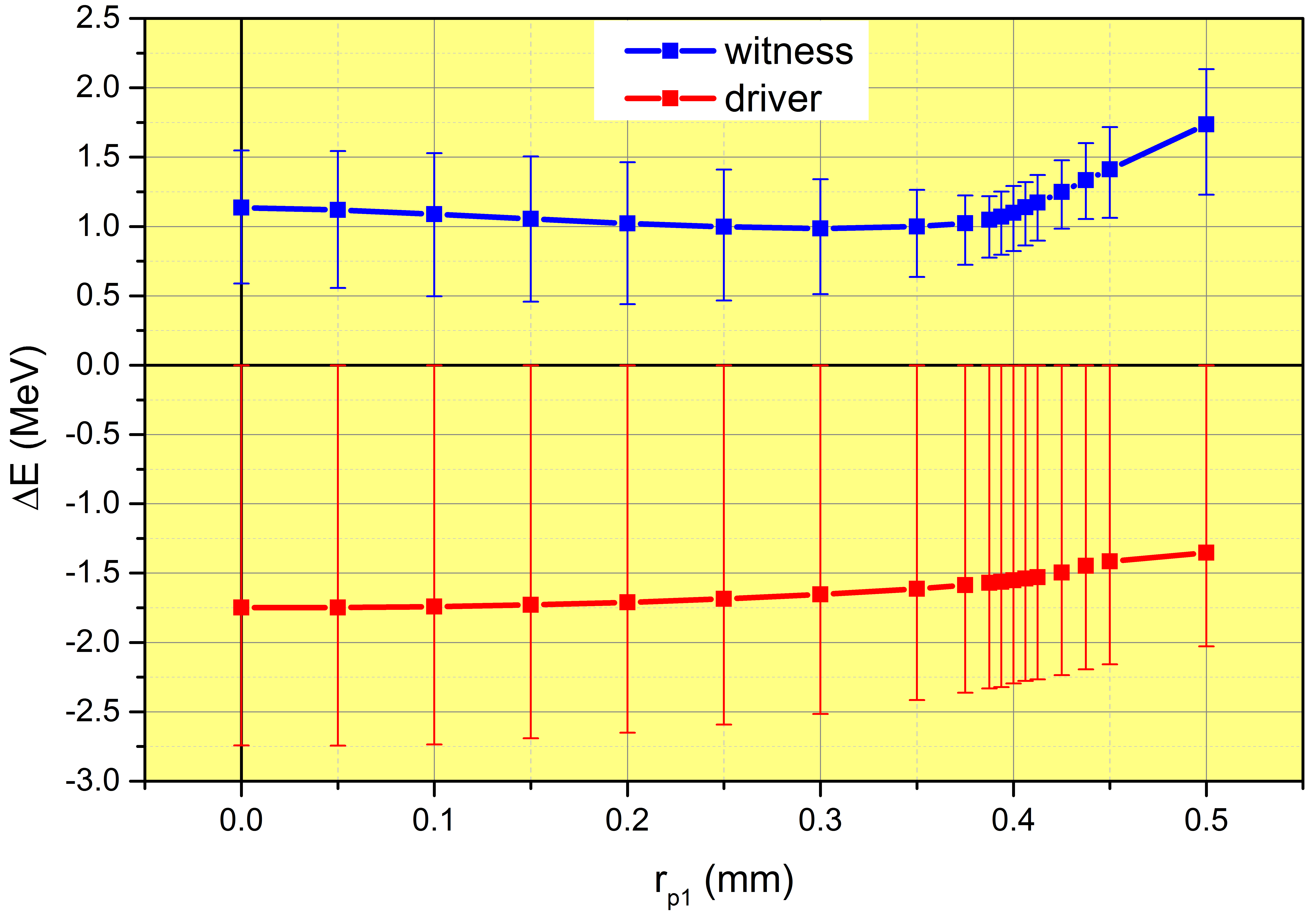}
  \caption{Energy gain of test bunch (blue curve) and energy loss of drive bunch (red curve) as functions of the vacuum channel radius $r_{p1}$   for the time $t = 26.69\;ps$ .}\label{Fig:08}
\end{figure}

Our present numerical experiments have also permitted us to estimate the influence of the plasma tube dimensions on the energy variations of the accelerated and drive bunches. It can be seen in \Figref{Fig:04} that in the absence of plasma at   (see \Figref{Fig:04}a), the test bunch is accelerated more intensively than with the presence of plasma in the drift channel (see \Figref{Fig:04}b, c and d). At the same time, at ${r_{p1}} = 0.4\;{\rm{mm}}$  (see \Figref{Fig:04}b), the accelerated electrons of the test bunch show a considerably smaller energy dispersion than at other   values shown in \Figref{Fig:04}. The above-stated regularities are obviously seen in \Figref{Fig:08}, which shows the energy changes of both the accelerated test bunch (blue curve) and the decelerated drive bunch (red curve) as functions of the vacuum channel radius   varying from $0$ to $0.5$ mm for the time $t = 26.69\;ps$. It can be also noticed from \Figref{Fig:08} that the smaller is the plasma tube thickness, the lower will be the energy losses and the electron energy dispersion of the driver bunch

\section{Conclusions}

The carried out numerical PIC simulation of wakefield excitation, as well as of self-consistent charged-particle dynamics in the plasma-dielectric slow-wave structure with a nonuniform transverse profile of the plasma density, has demonstrated that the vacuum channel in the plasma improves focusing of the accelerated bunch. There exists the optimum vacuum channel radius, at which the focusing proves to be the best.

The improvement in the accelerated bunch focusing in the presence of the vacuum channel is accompanied by a few decrease in the gradient of energy variation as compared with the full plasma filling of the drift channel.

It should be noted that the optimal acceleration takes place if the plasma is absent in the transport channel (traditional DWA), however, in that case no focusing of the test bunch occurs.

\begin{acknowledgments}
Work supported by NAS of Ukraine program "Perspective investigations on plasma physics, controlled thermonuclear fusion and plasma technologies", project P-1/63-2017 "Wakefield acceleration of electrons in multi-zone dielectric and plasma-dielectric structures".

Work was partially supported by the Ukrainian budget program "Support for the most important directions of scientific researches" ($K\Pi KBK$ 6541230).
\end{acknowledgments}

\bibliography{Bibliography}	
\end{document}